# Application of Origen2.1 in the decay photon spectrum calculation of spallation products[*]

Shuang Hong (洪爽)[1, 2]   Yong-Wei Yang (杨永伟)[1;1)]   Hu-Shan Xu(徐瑚珊)[1]
Hai-Yan Meng (孟海燕)[1]   Lu Zhang (张璐)[1]   Zhao-Qing Liu (刘照青)[1]
Yu-Cui Gao (高育翠)[1]   Kang Chen(陈康) [1]

[1] Department of Spallation Target, Institute of Modern Physics, Chinese Academy of Sciences, 509 Nanchang road, Lanzhou 730000, China

[2] University of Science and Technology of China, 96 Jinzhai road, Hefei 230026, China

**Abstract:** Origen2.1 is a widely used computer code for calculating the burnup, decay, and processing of radioactive materials. However, the nuclide library of Origen2.1 is used for existing reactors like pressurized water reactor, to calculate the photon spectrum released by the decay of spallation products, we have made specific libraries for the ADS tungsten spallation target, based on the results given by a Monte Carlo code: FLUKA. All the data used to make the Origen2.1 libraries is obtained from Nuclear structure & decay Data (NuDat2.6). The accumulated activity of spallation products and the contribution of nuclides to photon emission are given in this paper.

## 1   Introduction

Accelerator Driven subcritical System (ADS), consisting of proton accelerator, spallation target and subcritical reactor, is a widely recommended way to transmute the High-Level radioactive Wastes (HLW) produced by nuclear reactors [1]. To provide external neutrons for the subcritical reactor, the spallation target needs to be irradiated by high energy proton for a long term, which will lead to the accumulation of a large amount of radioactive products. So shielding design after the decommissioning of the spallation products is of necessity. Among all the radioactive shielding, gamma ray shielding is the most important one due to its strong penetrability. Here we use Origen2.1 to calculate the decay photon spectrum of spallation products to provide some reference data for the shielding of decommissioned spallation target.

FLUKA is a widely used Monte-Carlo code for simulating interaction and transport of hadrons, heavy ions and electromagnetic particles from thermal neutron energies (100eV) to cosmic ray energies (TeV) in whichever material [2]. In ADS studies, FLUKA is used to calculate the spallation neutron spectrum, accumulation of residual nucleus and design of the proton accelerator. It provides a card called RESNUCLE to calculate the spallation products. For low energy (<5GeV) projectiles like neutron, proton and π, a hadronic model of FLUKA, called PEANUT, is used to simulate the Pre-equilibrium stage. As for De-excitation stage, combination of models like Own Evaporation (or GEM), Fission, Multifragmentation and Fermi breakup are used to model the spallation reactions [3, 4]. Here we use FLUKA to calculate the accumulated activity of spallation products and provide it to Origen2.1 to calculate the decay photon spectrum.

[*] Supported by Strategic Priority Research Program of Chinese Academy of Sciences (XDA03030102)
1) Email: yangyongwei@impcas.ac.cn

Origen2.1 is a versatile point-depletion and radioactive-decay computer code for use in simulating nuclear fuel cycles and calculating the nuclide compositions and characteristics of materials contained therein [5]. It was developed at the Oak Ridge National Laboratory (ORNL) and distributed worldwide beginning in the early 1970s. With the features like a free-format input and various corresponding database, Origen2.1 is able to simulate varieties of fuel cycle flow sheets, the burnup of specific nuclides, decay photons released by nuclides are included [6]. The databases employed by Origen2.1 includes the decay, cross-section, fission product yield, and photon emission database. Since the Origen2.1 is used for existing reactors, including pressurized water reactors, boiling water reactors, liquid-metal fast breeder reactors, and Canada deuterium uranium reactors, it is not appropriate to apply Origen2.1 to calculate the decay photon spectrum of spallation products directly. So we have made specific libraries for a tungsten spallation target, containing the information of decay and photon production rates.

## 2  Calculation Method & Model

The calculation flow can be described as follows: firstly, we build a calculation model of spallation target by FLUKA, and calculate the accumulated activity of spallation products; secondly, we make the decay and photon libraries of relative isotopes, based on the results given by FLUKA; finally, we put the accumulated mass (in unit of grams) yield and the libraries together to Origen2.1 input and get the 18 energy group photon spectrum. Fig. 1 gives an intuitive view of the calculation flow.

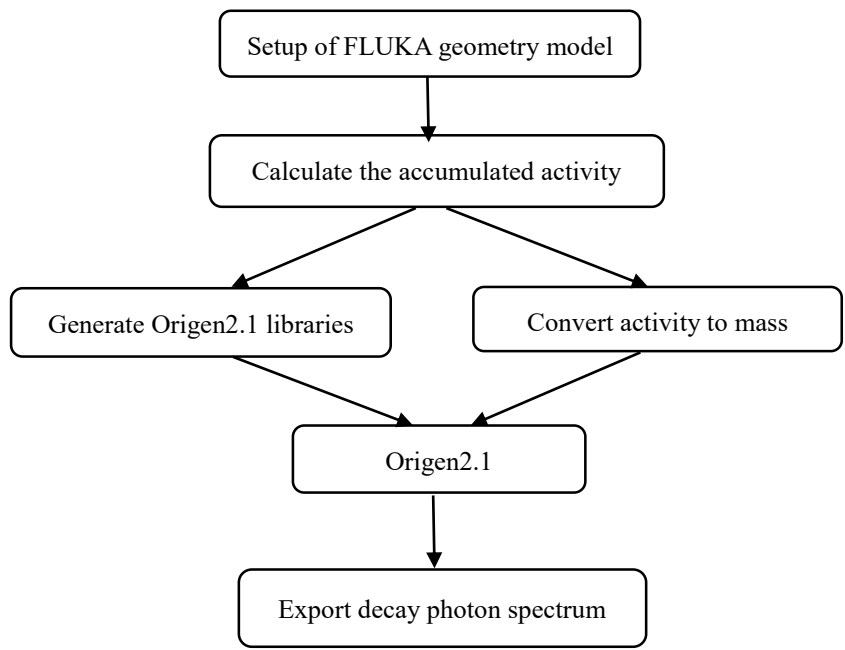

Fig.1. Calculation flow of decay photon spectrum

## 2.1 Calculation of spallation products

To generate the Origen2.1 libraries, we need to know the details of the spallation products. The geometry structure of spallation target is shown in Fig. 2. The target is a cylinder with a proton beam pipe inside it, protons generated from a linear accelerator will bombard with spallation material in a depth of 80cm. The thickness of beam and target tubes are 0.5cm and 1cm. The spallation target consists of tungsten, ion and nickel and the pipe wall material is T91.

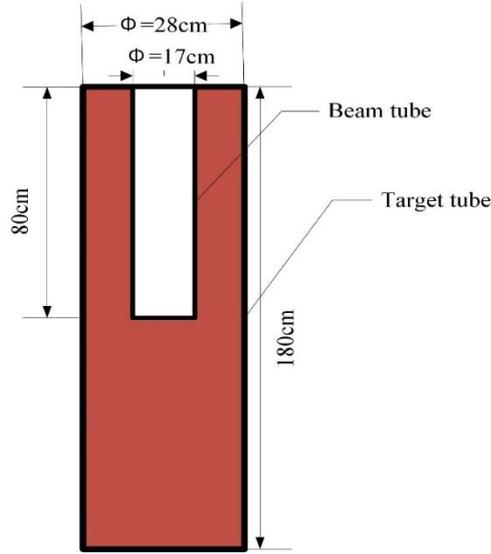

Fig.2. Geometry structure of spallation target

Here, we calculate the accumulated activities of the radioactive products combining the RADDECAY, RESUCLE, DCYTIMES, DCYCSCORE, and IRRPROFI card of FLUKA. The new evaporation model is activated to ensure a better quality of the results. Products scored here include spallation products (all interactions except those induced by neutrons below the threshold for multi-group treatment) and low-energy neutron products. The main parameters of the calculation model are given in Table 1. The preliminary particles here is set to 1E7.

Table 1. Calculation parameters

| Parameter | Material/Value |
|---|---|
| Energy of the Protons | 250 MeV |
| Current of the Accelerator | 10 mA |
| Proton beam distribution | Uniform |
| Radius of beam spot | 5 cm |
| Spallation target form | Granular |
| Spallation material | Tungsten93-Ion4.5-Nickel2.5 (Mass fraction %) |
| Equivalent density | 9.42g/cm$^3$ |
| Irradiation time | 5 years |

Since Origen2.1 input requires the mass yield of nuclides, the accumulated activity can be easily transferred to mass yield by equation listed below：

$$Y_i = \frac{A_i M_i}{\lambda_i N_A}. \qquad (1)$$

where $Y_i$ is the mass yield of nuclide $i$, $A_i$ is the activity given by FLUKA, $M_i$ is mass number, $\lambda_i$ is the radioactive decay constant and $N_A$ is Avogadro's constant.

**2.2 Generation of Origen2.1 libraries**

Origen2.1 external libraries consist of decay library, photon release library and cross-section library. The cross-section library is used to calculate the transmutation of wastes, since we neglect the transmutation of products by neutron, it is not necessary when calculating the decay photon spectrum. Information like half-life, decay branching ratio, average recovery energy and inhalation limits is included in the decay library. At present, we added half-life, decay branching ratio in the decay library, other values here are set to zeros. For the convenient of calculation, photon energies in Origen2.1 are divided into 18 groups. Table 2 gives the concrete energy group structures of photon library.

Table 2.  Origen2.1 photon library structures

| Group | Lower boundary(MeV) | Upper boundary(MeV) | Average energy(MeV) |
|---|---|---|---|
| 1 | 0.0000E-02 | 2.0000E-02 | 1.0000E-02 |
| 2 | 2.0000E-02 | 3.0000E-02 | 2.5000E-02 |
| 3 | 3.0000E-02 | 4.0000E-02 | 3.7500E-02 |
| 4 | 4.0000E-02 | 7.0000E-02 | 5.7500E-02 |
| 5 | 7.0000E-02 | 1.0000E-01 | 8.5000E-02 |
| 6 | 1.0000E-01 | 1.5000E-01 | 1.2500E-01 |
| 7 | 1.5000E-01 | 3.0000E-01 | 2.2500E-01 |
| 8 | 3.0000E-01 | 4.5000E-01 | 3.7500E-01 |
| 9 | 4.5000E-01 | 7.0000E-01 | 5.7500E-01 |
| 10 | 7.0000E-01 | 1.0000E 00 | 8.5000E-01 |
| 11 | 1.0000E 00 | 1.5000E 00 | 1.2500E 00 |
| 12 | 1.5000E 00 | 2.0000E 00 | 1.7500E 00 |
| 13 | 2.0000E 00 | 2.5000E 00 | 2.2500E 00 |
| 14 | 2.5000E 00 | 3.0000E 00 | 2.7500E 00 |
| 15 | 3.0000E 00 | 4.0000E 00 | 3.5000E 00 |
| 16 | 4.0000E 00 | 6.0000E 00 | 5.0000E 00 |
| 17 | 6.0000E 00 | 8.0000E 00 | 7.0000E 00 |
| 18 | 8.0000E 00 | 1.0000E 01 | 9.5000E 00 |

To generate the decay and photon release libraries, we get data from Nuclear Structure & decay Data (NuDat2.6), and reprocess these data into corresponding format requested by Origen2.1. Since the photon release rate given by NuDat2.6 is discrete in energy, we count the release rate falling in each region listed above, and obtain the photon library.

**2.3 Calculation of decay photon spectrum**

When calculating the decay photon spectrum, the accumulated nucleus of specific nuclide is necessary. In Origen2.1, the amount of nuclide $N_i$ changes as a function of time is described by a nonhomogeneous first-order ordinary differential equation as follows [6]:

$$\frac{dN_i}{dt} = \sum_{j=1}^{N} l_{ij}\lambda_j N_j + \phi \sum_{k=1}^{N} f_{ik}\sigma_k N_k - (\lambda_i + \phi\sigma_i + r_i)N_i + F_i, i = 1......N \quad (2)$$

Where

$N_i$ = atom density of nuclide $i$

$N$ = number of nuclides

$l_{ij}$ = fraction of radioactive disintegration by other nuclides, which leads to formation of nuclide $i$

$\lambda_i$ = radioactive decay constant

$\Phi$ = position and energy-averaged neutron flux, in calculation it is set to 0

$f_{ik}$ = fraction of neutron absorption by other nuclides, which leads to formation of nuclide $i$

$\sigma_k$ = spectrum-averaged neutron absorption cross section of nuclide $k$

$r_i$ = continuous removal rate of nuclide I from the system

$F_i$ = continuous feed rate of nuclides $i$

By solving these $N$ equations, Origen2.1 gives the amounts of each nuclide at the end of each time step set in DEC card. And photon release rate by nuclide $i$ thus can be calculated from equation:

$$P_i = g_i \lambda_i N_i. \quad (3)$$

where $P_i$ is the photon release rate of group $i$, $g_i$ is the photon released per decay, which is given in Origen2.1 photon library, $\lambda_i$ and $N_i$ are described above.

After getting the spallation yield and Origen2.1 libraries, the 18 energy group photon release data can be easily obtained using the DEC card.

## 3 Results and Discussion

### 3.1 Accumulated activity of spallation products

Based on the geometry and calculation parameters mentioned above, the accumulated activity of spallation products can be calculated by FLUKA. The spallation material consists of natural tungsten and iron/nickel. Table 3 gives the main radioactive isotopes produced in the target region. Here, the nuclide ID is defined as 1000A+Z, where A is the mass number and Z is the atomic number. It should mention that the products scored here include all spallation products and low-energy neutron products. The activity of tungsten products is greater than iron/nickel by an order of one magnitude.

Table 3. Main radioactive products of spallation materials

| Spallation material | Main products(Nuclide ID) | Activity (Bq) | Error (%)[1] | Half-life (day) |
|---|---|---|---|---|
| $W^{nat}$ (93%)[2] | 74185 | 4.97E+15 | 0.1338 | 75.10 |
| | 74181 | 4.09E+15 | 0.1176 | 121.20 |
| | 74187 | 2.81E+15 | 6.43E-02 | 0.99 |
| | 74179 | 2.52E+15 | 0.1052 | 0.03 |
| | 73179 | 2.47E+15 | 0.1185 | 664.30 |
| $Fe^{nat}$(4.5%)/$Ni^{nat}$(2.5%) | 26055 | 4.40E+14 | 0.3758 | 999.01 |
| | 27058 | 3.63E+14 | 0.2497 | 70.86 |
| | 27057 | 2.75E+14 | 0.4293 | 271.74 |
| | 25054 | 2.33E+14 | 0.3785 | 312.03 |
| | 24051 | 1.55E+14 | 0.5208 | 27.70 |

---

[1] Statistical error of activity in percentage given by FLUKA, that is $\sigma/N^{1/2}$, σ stands for standard deviation, N stands for number of event.

[2] Mass fraction, superscript "nat" stands for natural.

We can see that the main radiation in the products is induced by the isotopes of spallation materials in Fig.3. It shows there are three peaks in the distribution of radioactive products, which are induced by light evaporation fragments, iron/nickel and tungsten. The highest activity appears at atomic number equals to 74, and reaches nearly $5\times10^{15}$Bq, the peak caused by iron/nickel is approximately $5\times10^{14}$Bq.

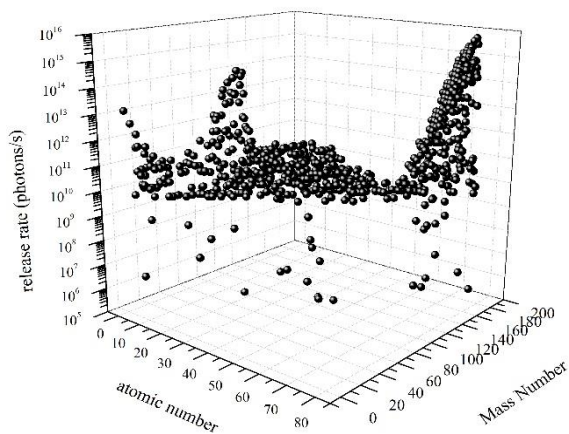

Fig.3. Distribution of radioactive products

## 3.2 Decay photon spectrum

The decay library we have made includes about 370 kinds of spallation products and the photon library contains about 260 kinds of nuclides, which have the activity higher than $1\times10^{12}$Bq. After generating the Origen2.1 libraries, we calculate the decay photon release rate at different cooling times. The result is shown in Table 4. The main photon emitters are isotopes of W, Ta, Lu, Re, Yb and Hf, most of which have short half-life and thus contribute little to photon emission after a comparative long cooling time. For nuclides like $^{44}$Ti, $^{90}$Y, $^{157}$Tb, $^{94}$Nb, and their daughters, they contribute more than 90% to photon emission after a cooling time of 100 years, and the total photon release rate decreases about 5 orders of magnitude. The total gamma power is $3.71\times10^{3}$W, which is relatively low and thus is negligible.

Table 4. 18 energy group decay photon spectrum at different cooling time

| Emean(MeV)\Cooling time | 0 s | 1 day | 100 days | 1 year | 100 years |
| --- | --- | --- | --- | --- | --- |
| 1.00E-02 | 1.75E+16 | 1.00E+16 | 2.29E+15 | 1.17E+15 | 1.17E+10 |
| 2.50E-02 | 1.51E+14 | 1.24E+14 | 1.02E+14 | 7.62E+13 | 4.56E+07 |
| 3.75E-02 | 6.15E+14 | 1.14E+14 | 5.41E+12 | 2.58E+12 | 1.46E+09 |
| 5.75E-02 | 3.37E+16 | 1.92E+16 | 5.21E+15 | 2.70E+15 | 3.12E+08 |
| 8.50E-02 | 1.64E+15 | 6.95E+14 | 2.05E+14 | 1.32E+14 | 2.91E+07 |
| 1.25E-01 | 4.09E+15 | 1.57E+15 | 4.34E+14 | 2.07E+14 | 6.39E+07 |
| 2.25E-01 | 5.86E+15 | 1.91E+15 | 6.96E+14 | 2.72E+14 | 2.58E+07 |
| 3.75E-01 | 3.02E+15 | 8.43E+14 | 1.84E+14 | 6.27E+13 | 1.85E+07 |
| 5.75E-01 | 7.16E+15 | 2.21E+15 | 3.40E+14 | 1.06E+14 | 9.48E+10 |
| 8.50E-01 | 4.16E+15 | 1.91E+15 | 7.33E+14 | 3.76E+14 | 1.34E+08 |
| 1.25E+00 | 4.57E+15 | 1.72E+15 | 5.80E+14 | 3.41E+14 | 5.07E+10 |
| 1.75E+00 | 1.62E+15 | 3.48E+14 | 5.29E+13 | 3.40E+13 | 1.09E+05 |
| 2.25E+00 | 2.24E+14 | 9.58E+13 | 6.09E+12 | 5.62E+11 | 2.06E+06 |
| 2.75E+00 | 2.36E+14 | 9.02E+13 | 8.84E+12 | 8.17E+11 | 5.62E+07 |
| 3.50E+00 | 1.54E+14 | 3.27E+13 | 7.90E+12 | 7.30E+11 | 7.02E+05 |
| 5.00E+00 | 2.88E+10 | 1.48E+09 | 6.85E+03 | 1.72E-07 | 1.72E-07 |
| 7.00E+00 | 9.77E+07 | 1.11E-08 | 1.11E-08 | 1.11E-08 | 1.11E-08 |
| 9.50E+00 | 7.05E-10 | 7.05E-10 | 7.05E-10 | 7.05E-10 | 7.05E-10 |
| TOTAL | 8.48E+16 | 4.09E+16 | 1.09E+16 | 5.48E+15 | 1.59E+11 |
| γPower(W) | 3.71E+03 | 1.35E+03 | 3.73E+02 | 1.87E+02 | 1.90E-02 |

The decay photon spectrum calculated by Origen2.1 at different cooling time is shown in Fig.4. It states that most photons release by decay of spallation products are in range of 0-1MeV. The high energy group photon release rate is negligible. After a cooling time of 1 year and 100 years, the photon release rate will decrease about 10 and $10^6$ times.

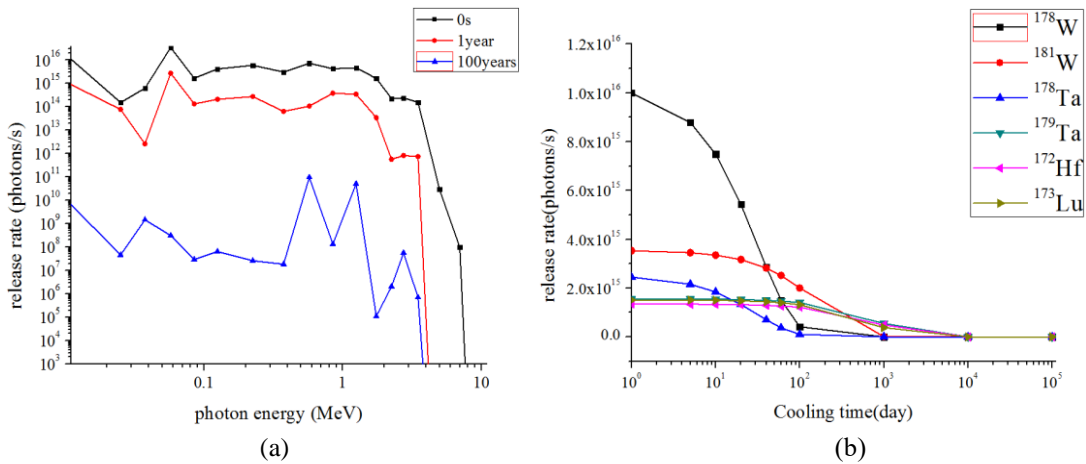

(a)  (b)

Fig. 4.　(a) Decay photon spectrum at different cooling time (b) Changing of total photon release rate of main photon emitters with respect to cooling time

## 4　Conclusion

In our work, specific Origen2.1 libraries for a tungsten spallation target have been made for further relative calculation. The accumulated activity of products induced by 250MeV, 10mA proton beam is calculated by a Monte-Carlo code: FLUKA, and the distribution of activity on atomic and mass number are given above. The main decay photon release rate calculated by Origen2.1 is at a magnitude of $10^{15}$, and the energy is concentrated in low energy region, that is 0~1MeV (account for 91.86% of the total).

The main photon emitters in photon group 1 is given in Table 5. For the whole 18 group photons, the result shows that the main photon emitters like $^{177}$W, $^{177}$Ta and $^{73}$As are of relatively short half-life, and will decrease rapidly, which could be seen in Fig. 4(b) intuitively. After a long cooling time, the main emitters would be $^{44}$Ti, $^{90}$Y, $^{157}$Tb, $^{94}$Nb, and their daughters.

Table 5. Photon release rate of main photon emitters in energy group 1

| Photon emitter\Cooling time | 0 s | 1 day | 100 days | 1 year | 100 years |
|---|---|---|---|---|---|
| $^{178}$W | 4.69E+15 | 4.54E+15 | 1.89E+14 | 3.81E+10 | 0.00E+00 |
| $^{73}$As | 1.78E+15 | 0.00E+00 | 0.00E+00 | 0.00E+00 | 0.00E+00 |
| $^{177}$W | 1.02E+15 | 5.55E+11 | 0.00E+00 | 0.00E+00 | 0.00E+00 |
| $^{181}$W | 7.74E+14 | 7.70E+14 | 4.37E+14 | 9.59E+13 | 0.00E+00 |
| $^{179}$W | 7.30E+14 | 2.03E+03 | 0.00E+00 | 0.00E+00 | 0.00E+00 |
| $^{176}$Ta | 6.44E+14 | 1.13E+14 | 0.00E+00 | 0.00E+00 | 0.00E+00 |
| $^{175}$Ta | 5.25E+14 | 1.13E+14 | 0.00E+00 | 0.00E+00 | 0.00E+00 |
| $^{174}$Lu | 5.21E+14 | 4.22E+14 | 4.19E+05 | 0.00E+00 | 0.00E+00 |
| $^{178}$Ta | 5.19E+14 | 4.70E+14 | 1.96E+13 | 3.94E+09 | 0.00E+00 |
| $^{177}$Ta | 5.15E+14 | 3.97E+14 | 9.16E+01 | 0.00E+00 | 0.00E+00 |
| $^{179}$Ta | 4.99E+14 | 4.98E+14 | 4.49E+14 | 3.41E+14 | 1.44E-02 |
| $^{172}$Hf | 4.07E+14 | 4.06E+14 | 3.67E+14 | 2.81E+14 | 3.25E-02 |
| $^{174}$Ta | 3.85E+14 | 3.11E+08 | 0.00E+00 | 0.00E+00 | 0.00E+00 |
| $^{173}$Ta | 3.57E+14 | 1.84E+12 | 0.00E+00 | 0.00E+00 | 0.00E+00 |
| $^{171}$Lu | 3.20E+14 | 3.08E+14 | 7.57E+10 | 1.55E+01 | 0.00E+00 |
| $^{181}$Re | 3.18E+14 | 1.38E+14 | 0.00E+00 | 0.00E+00 | 0.00E+00 |
| $^{173}$Lu | 3.07E+14 | 3.07E+14 | 2.68E+14 | 1.85E+14 | 3.27E-08 |
| $^{180}$Re | 2.54E+14 | 3.15E-10 | 0.00E+00 | 0.00E+00 | 0.00E+00 |
| $^{179}$Re | 2.53E+14 | 0.00E+00 | 0.00E+00 | 0.00E+00 | 0.00E+00 |
| $^{44}$Sc | 7.33E+10 | 2.41E+09 | 1.31E+09 | 1.30E+09 | 4.14E+08 |
| $^{44}$Ti | 3.01E+10 | 3.01E+10 | 3.00E+10 | 2.97E+10 | 9.48E+09 |
| $^{90}$Y | 2.49E+10 | 1.96E+10 | 1.79E+09 | 1.76E+09 | 1.67E+08 |
| $^{157}$Tb | 2.24E+09 | 2.24E+09 | 2.23E+09 | 2.23E+09 | 1.41E+09 |

After a cooling time of 100 years, the total photon release rate will decrease from $8.48\times10^{16}$ to $1.59\times10^{11}$ photons/s. And thus there is no need to worry about the long period disposal for the spallation target from the aspect of gamma ray shielding.